\documentclass[amsmath,amssymb,article,preprints,article,accept,moreauthors,pdftex,10pt,a4paper]{revtex4-1}

\usepackage[utf8x]{inputenc}
\usepackage{ucs}

\usepackage{booktabs}

\usepackage{verbatim}
\usepackage{amsfonts}
\usepackage{dcolumn}
\usepackage{bm}
\usepackage[usenames]{color}
\usepackage{multirow}
\usepackage{graphicx}
\usepackage{hyperref}
\usepackage{subfig}
\usepackage{booktabs}
\usepackage{xcolor}
\usepackage[normalem]{ulem}

\usepackage{ucs}
\usepackage{amssymb}
\usepackage{makeidx}
\usepackage{diagbox}
\usepackage{tabularx}

\usepackage{amsthm}

\usepackage{float} 
\usepackage{wrapfig} 

\usepackage{upgreek} 
\usepackage{cancel} 
\usepackage{mathdots} 
\usepackage{mathrsfs} 
\graphicspath{{figures/}} 

\begin{document}	
\title{Cr\'amer-Rao complexity of the two-dimensional confined hydrogen}
%

\author{C. R. Esta\~{n}\'on}
\email[]{carloscbiuam1@gmail.com}
\affiliation{%
	Departamento de  F\'isica, Universidad Aut\'onoma Metropolitana-Iztapalapa,
	Av. San Rafael Atlixco $186$, Col. Vicentina, CP $09340$ CDMX, M\'exico
}%

\author{N. Aquino }
\email[]{naa@xanum.uam.mx}
\affiliation{%
	Departamento de  F\'isica, Universidad Aut\'onoma Metropolitana-Iztapalapa,
	Av. San Rafael Atlixco $186$, Col. Vicentina, CP $09340$ CDMX, M\'exico
}%


\author{David Puertas-Centeno}

\email{david.puertas@urjc.es}
\affiliation{%
	Departamento de Matem\'atica Aplicada, Universidad Rey Juan Carlos, 28933 Madrid, Spain
}%

\author{Jes\'us S. Dehesa}
\email[]{dehesa@ugr.es}
\affiliation{%
	Departamento de F\'{\i}sica At\'{o}mica, Molecular y Nuclear, Universidad de Granada, 18071 Granada, Spain and
	Instituto Carlos I de F\'{\i}sica Te\'orica y Computacional, Universidad de Granada, 18071 Granada, Spain
}%
%


%

\date{\today}

\begin{abstract}
	The internal disorder of the two-dimensional confined  hydrogenic atom is numerically studied in terms of the confinement radius for the 1\textit{s}, 2\textit{s}, 2\textit{p} and 3\textit{d} quantum states by means of the statistical Cr\'amer-Rao complexity measure. First, the confinement dependence of the variance and the Fisher information of the position and momentum spreading of its electron distribution are computed and discussed. Then, the Cr\'amer-Rao complexity measure (which quantifies the combined balance of the charge concentration around the mean value and the gradient content of the electron distribution) is investigated in position and momentum spaces. We found that confinement does distinguish complexity of the system for all quantum states by means of this two component measure.
\end{abstract}

\maketitle

\section{Introduction}
The internal disorder of $d$-dimensional quantum-mechanical non-relativistic systems is
conditioned by the spatial spreading/complexity of their Schr\"odinger single-particle
probability density $\rho(\vec{r}), \vec{r} \in \mathbb R_d$ \cite{Herschbach1993,Sen2011}. To best quantify it, three composite information-theoretic measures (the Fisher-Shannon, Lopezruiz-Mancini-Calvet (LMC) and Cr\'amer-Rao complexities) \cite{Dehesa2011,Angulo2011} and various generalizations (e.g., the complexities of Fisher-R\'enyi and LMC-R\'enyi type) have been proposed beyond the single dispersion (the statistical variance) and information-theoretic measures (the Fisher information and the Shannon entropy) and their extensions (e.g., the entropies of R\'enyi and Tsallis type) \cite{Dehesa2010}. The latter measures are given as $V[\rho] = \left\langle r^2 \right\rangle - \left\langle r\right\rangle^2$, $r \equiv |\vec{r}|$, for the variance 
and 
\begin{equation}
F\left[\rho \right]:= \int_{\mathbb R_d} \frac{\left| \vec{\nabla}_d\,\rho (\vec{r})
	\right|^2 }{\rho(\vec{r})} d\vec{r}, \quad \quad S\left[ \rho \right]:= - \int_{\mathbb R_d} \rho (\vec{r})  \log \rho(\vec{r}) d\vec{r}
\end{equation}
for the Fisher and Shannon infromation-theoretic entropies \cite{Fisher1925,Shannon1948}, respectively. The symbol $\bar{\nabla}_d$ denotes the $d$-dimensional gradient operator. 
These measures quantify a single facet of the density $\rho(\vec{r})$ of local (Fisher) or global (variance, Shannon) character, such as the concentration around the mean value (variance), the gradient content (Fisher information) and the total extent (Shannon entropy) of the density. They have been shown to be very useful in numerous scientific areas, particularly to identify and characterize many atomic, molecular and chemical phenomena such as e.g., correlation properties, level avoided crossings of atoms in external electromagnetic fields, and transition states and other stationary points in chemical reactions \cite{Gonzalez2003,Gonzalez2005,Esquivel2015,Esquivel2016,Lopezrosa2016}.

The composite information-theoretic measures have been recently shown to be most appropriate to describe the intrinsic complexity of the quantum systems and to distinguish among
their rich three-dimensional geometries, mainly because they jointly grasp different facets of their internal disorder. This is basically because (i) they are dimensionless, (ii) they are invariant under replication (LMC), translation and scaling transformations, and (iii) they have, under certain mathematical conditions, minimal values for both extreme cases: the completely ordered systems
(e.g. a Dirac delta distribution and a perfect crystal in one and three
dimensions) and the totally disordered systems (e.g., an uniform or highly flat
distribution and an ideal gas in one and three dimensions). The basic composite information-theoretic measures have the following two-ingredient expressions 
\begin{equation} \label{eq:lmc}
C_{LMC}[\rho]= e^{S_\rho}\times\int_{\mathbb R_d}[\rho(\vec r)]^2\,d\vec r,\quad \quad C_{\text{FS}}[\rho] = F[\rho] \times \frac{1}{2\pi e} e^{\frac{2}{d} S\left[\rho \right]}
\end{equation}
for the LMC (Lopezruiz-Mancini-Calvet) \cite{Lopez1995,Catalan2002} and the Fisher-Shannon complexities $C_{\text{FS}}[\rho]$ \cite{Romera2004,Angulo2008}, respectively, and
\begin{equation} \label{eq:cramer}
C_{\text{CR}}[\rho] =F[\rho]\times V[\rho]
\end{equation}
for the Cr\'amer-Rao complexity \cite{Angulo2008a}.
Note that the LMC complexity measures the density non-uniformity through  the combined effect of its total spreading and  average height, while the Fisher-Shannon complexity grasps the oscillatory nature of the density together with its total extent in the configuration space, and the Cr\'amer-Rao quantity takes into account the gradient content of the density jointly with its concentration around the centroid. The Cr\'amer-Rao complexity measure occupies a special position among the intrinsic two-product complexity quantifiers of the internal disorder of the quantum systems. Indeed, it is the ONLY one which measures the localization of the probability density around its nodes (as given by the Fisher information) jointly with the concentration around the mean (as given by the variance). The variance measures an average of distances of outcomes of the probability distribution from the mean. Although the Shannon entropy, the disequilibrium, the R\'enyi entropies (which are components of the other two-product complexities) and the variance are measures of dispersion and uncertainty, the lack of a simple relationship between orderings of a distribution by these entropy-like and variance measures emanates from quite substantial and subtle differences. Both entropy-like and variance measures reflect concentration but their respective metrics for concentration are different. Unlike variance which measures concentration around the mean, entropy-like measures quantify  diffuseness of the density irrespective of the location(s) of concentration.\\

The study of the internal disorder for the confined multidimensional quantum systems by means of entropy- and complexity-like measures is an area which has been increased momentously \textit{per se} (the confinement effects on the charge and momentum distributions of the quantum systems are not yet well established) and because of the growing number of relevant applications in numerous fields of science and engineering, from surface chemistry to quantum technologies where e.g. confined atoms have been recently suggested as building elements of qubits. This is illustrated by the various reviews/states-of-the-art appeared in the last few years \cite{Sabin2009,Sen2014,Inglesfield2015,LeyKoo2018}.\\  

Here we are interested in the electronic complexity of the spherically-confined two-dimensional hydrogen atom (in short, 2D-HA) both in position and momentum spaces. This atom is the prototype which has been used to interpret numerous phenomena and systems in surface chemistry \cite{Cox1985,Jakubith1985,Beta1985}, semiconductors (see e.g., \cite{Li2007}), quantum dots \cite{Harrison2005,Munjal2018}, atoms and molecules embedded in nanocavities (e.g., fullerenes, helium droplets,...) \cite{Sabin2009,Sen2014,LeyKoo2018,Al-Hashimi2012,Aquino2014,Rojas2018}, dilute bosonic and fermionic systems in magnetic traps of extremely low temperatures \cite{Gleisberg2000,Anglin2002,DeMarco1999} and a variety of quantum-information elements \cite{Lukin2018,Yoder2019}. Up until now, contrary to the stationary states of the confined 3D-HA where both the (energy-dependent) spectroscopic and the (eigenfunction-dependent) information-theoretic properties have received much attention \cite{Sen2005,Aquino2013, Nascimento2018,Jiao2017,Mukherjee2018,Mukherjee2018b,Mukherjee2018a,Munjal2018,Garza2019,Wu2020,Majumdar2017}, the knowledge of these properties for the confined 2D-HA is quite scarce \cite{Yang1991,Aquino1998,Aquino1998a,Aquino2005,Chaos2005}. Just recently, the authors have determined \cite{Estanon2020} the entropy-like (Shannon, Fisher) and complexity-like (Fisher-Shannon, LMC) measures for a few low-lying stationary states of the confined 2D-HA.  The aim of this work is to extend this informational approach by means of the calculation  of the confinement dependence of the variance and the Cr\'amer-Rao complexity measure for the $1\textit{s}, 2\textit{s}, 2\textit{p}$ and $3\textit{d}$ quantum states of the 2D-HA in the two conjugated spaces.\\

The structure of this work is the following. In Section \ref{sec:freeHA} we first analytically discuss the eigenvalue free (unconfined) $d$-dimensional hydrogen problem from an informational point of view, with application to the 1\textit{s}, 2\textit{s}, 2\textit{p} and 3\textit{d} states of the free two-dimensional case. In Section \ref{sec:tres} the computational method to solve the eigenvalue confined two-dimensional hydrogen problem is described and the probability densities which characterize the stationary states of this system are given in both position and momentum spaces. In Section \ref{sec:cuatro}, we calculate and discuss the variance, the Fisher information and the Cr\'amer-Rao complexity of these densities in both conjugated spaces for various stationary states of the confined 2D-HA. Therein it is explicitly shown how the values of these quantities go towards the corresponding free constant values analytically found in Table I of Section \ref{sec:freeHA}, what is a further checkin of our results. Finally, some concluding remarks are made.\\

\section{The eigenvalue $d$-dimensional hydrogen problem: An informational approach}
\label{sec:freeHA}
In this section we briefly describe the analytical solution of the non-relativistic eigenvalue problem for the free (i.e., unconfined) $d$-dimensional hydrogen atom in both position and momentum spaces. The main purpose of this section is to analytically obtain the expressions of the variance and the Fisher information of the free two-dimensional hydrogen atom, and their corresponding products which give the Cr\'amer-Rao complexity measures in the two conjugated spaces as shown in Table I for various low-lying states. They will be used in Section \ref{sec:cuatro} to check the validity of our confined complexity results  when confinement gets weaker and weaker. The resulting electron probability densities are used to determine the dispersion and Fisher information measures of the $(n\textit{s})$ and circular states of this system in a rigorous way, with applications to the $1\textit{s}, 2\textit{s}, 2\textit{p}$ and $3\textit{d}$ stationary states of the free 2D-HA. The Schr\"{o}dinger equation of the free (i.e., unconfined) $d$-dimensional hydrogenic system has the form 
\begin{equation}\label{schrodinger}
\left( -\frac{1}{2} \vec{\nabla}^{2}_{d} + \mathcal{V}(r)\right) \Psi \left( \vec{r} \right) = E \Psi \left(\vec{r} \right),
\end{equation}
in atomic units, where $\vec{r}  =  (r,\theta_1,\theta_2,\ldots,\theta_{D-1})$ in hyperspherical units and $r \equiv |\vec{r}|
\in [0 , \: +\infty)$. The symbols  $\vec{\nabla}_{d}$ and $\mathcal{V}(r)$ denote the $d$-dimensional gradient operator and the Coulomb potential $\mathcal{V}(\vec{r}) =  \frac{1}{r}$, respectively. It has been shown \cite{Yanez1994,Dehesa2010} that the wavefunctions of this system are characterized by the energies 
\begin{equation} \label{eqI_cap1:energia}
E= -\frac{1}{2\eta^2},\hspace{0.5cm} \hspace{0.5cm} \eta=n+\frac{d-3}{2}; \hspace{5mm} n=1,2,3,...,
\end{equation}
and the associated eigenfunctions
\begin{equation}\label{eqI_cap1:FunOnda_P}
\Psi_{n,l, \left\lbrace \mu \right\rbrace }(\vec{r})=R_{n,l}(r)\times {\cal{Y}}_{l,\{\mu\}}(\Omega_{d-1}),
\end{equation}
where $(l,\left\lbrace \mu \right\rbrace)\equiv(l\equiv\mu_1,\mu_2,...,\mu_{d-1})$ denote the hyperquantum numbers associated to the angular variables $\Omega_{d-1}\equiv (\theta_1, \theta_2,...,\theta_{d-1})$, which may take all values consistent with the inequalities $l\equiv\mu_1\geq\mu_2\geq...\geq \left|\mu_{d-1} \right| \equiv \left|m\right|\geq 0$. The radial part of the eigenfunction is given by
\begin{equation}\label{eqI_cap1:Rnl}
R_{n,l}(r)=\left( \frac{\lambda^{-d}}{2 \eta}\right)^{1/2}   \left[\frac{\omega_{2L+1}(\tilde{r})}{\tilde{r}^{d-2}}\right]^{1/2}{\tilde{\cal{L}}}_{\eta-L-1}^{(2L+1)}(\tilde{r}),
\end{equation}
where $\lambda=\frac{\eta}{2}$,\,$\tilde{r}=\frac{r}{\lambda}$, $L$ is 
\begin{equation} \label{eqI_cap1:Lyr}
L=l+\frac{d-3}{2}, \hspace{0.5cm} l=0, 1, 2, \ldots
\end{equation}
and the symbol ${\tilde{\cal{L}}}^{(\alpha)}_{k}(x)$ denotes the orthonormal Laguerre polynomial of degree $k$ with respect to the weight $\omega_\alpha(x)=x^{\alpha} e^{-x}$ on the interval $\left[0,\infty \right) $. The angular part of the eigenfunction is given by the known hyperspherical harmonics ${\cal{Y}}_{l,\{\mu\}}(\Omega_{d-1})$ \cite{Yanez1994,Dehesa2010}.\\
Then, the position probability density for a generic $(n,l,\left\lbrace \mu \right\rbrace)\equiv(n,l\equiv\mu_1,\mu_2,...,\mu_{d-1})$ state of the free $d$-dimensional hydrogenic systems is
\begin{equation}\label{eqI_cap1:DenPos}
\rho(\vec{r}) =\left|\Psi_{n,l,\{\mu\}}\left( \vec{r} \right) \right|^2=R^{2}_{n,l} \left( r \right)\times \left|{\cal{Y}}_{l,\{\mu\}}(\Omega_{d-1})\right|^2.
\end{equation}
which is normalized so that $\int \rho(\vec{r}) d\vec{r}=1$.\\
And the probability density in momentum spaces $\gamma(\vec{p})$ is obtained by squaring the $d$-dimensional Fourier transform of the configuration eigenfunction, i.e., the momentum eigenfunction \cite{Dehesa2010}
\begin{equation}\label{eqI_cap1:FunOnda_M}
\tilde{\Psi}_{n,l,\{\mu\}}(\vec{p})={\cal{M}}_{n,l}(p)\times {\cal{Y}}_{l\{\mu\}}(\Omega_{d-1}),
\end{equation}
whose radial part is
\begin{equation}\label{eqI_cap1:Mnl}
{\cal{M}}_{n,l}(p)=\left(  \frac{\eta} {Z}\right) ^{d/2} (1+y)^{3/2} \left(\frac{1+y}{1-y} \right)^{\frac{d-2}{4}} \sqrt{\omega^{*}_{L+1} (y)} {\tilde{C}}^{L+1}_{\eta-L-1}(y).
\end{equation}
where $y \equiv (1-\eta^2 p^2)/(1+\eta^2 p^2)$ and the symbol ${\tilde{C}}^\alpha_m(x)$ denotes the orthonormal Gegenbauer polynomials with respect to the weight funciont $\omega^{*}_{\alpha}(x)=(1-x^2)^{\alpha-\frac{1}{2}}$ on the interval $\left[-1,+1 \right] $. Then, the momentum probability density for a generic $(n,l,\left\lbrace \mu \right\rbrace)\equiv(n,l\equiv\mu_1,\mu_2,...,\mu_{d-1})$ state of the free $d$-dimensional hydrogenic systems is
\begin{equation}\label{eqI_cap1:DenMom}
\gamma(\vec{p})=\left|\tilde{\Psi}_{n,l,\{\mu\}}\left( \vec{p} \right) \right|^2={\cal{M}}^2_{n,l}(p) \left[ {\cal{Y}}_{l\{\mu\}}(\Omega_{d-1})\right]^2,
\end{equation}
which is normalized so that $\int \gamma(\vec{p}) d\vec{p}=1$.\\
Let us highlight that all the dispersion, entropy-like and complexity-like which quantify the different facets of the electron delocalization and complexity of the free $d$-dimensional hydrogenic system in both position and momentum spaces can be obtained by calculating the corresponding expressions given in the previous section for the position and momentum densities of the system which we have just found. In particular, since
\begin{align}\nonumber
\left\langle r^\alpha \right\rangle & \equiv \int_{\mathbb R_d} r^\alpha \rho(\vec{r}) d\vec{r}=\int^{\infty}_{0}r^{\alpha+d-1} R^{2}_{nl} \left( r \right) dr\\\label{eqI_cap1:<ra>}
&=\frac{1}{2 \eta} \left( \frac{\eta}{2}\right)^\alpha  \int^{\infty}_{0} \omega_{2L+1}\left( \tilde{r}\right) \left[ {\tilde{\cal{L}}}^{(2L+1)}_{n_r} \left( \tilde{r}\right) \right]^2 \tilde{r}^{\alpha+1} d\tilde{r},
\end{align}
we have the values \cite{Dehesa2010}.
\begin{equation}\label{eqI_cap1:<r>}
\left\langle r \right\rangle=\frac{1}{2} \left[3 \eta^2-L(L+1) \right]; \quad \left\langle r^2 \right\rangle=\frac{1}{2}\eta^2 \left[5 \eta^2-3L(L+1)+1 \right],
\end{equation}
so that the variance $V[\rho] = \left\langle r^2 \right\rangle - \left\langle r\right\rangle^2$ has the value
\begin{equation}\label{eqI_cap1:varianza_rho}
V\left[ \rho \right]=\frac{1}{4}[\eta^2 (\eta^2+2)-L^2 (L+1)^2].
\end{equation}
In addition, one can obtain the values \cite{Dehesa2006,Romera2006}
\begin{align}
F\left[ \rho \right]&=4 \left\langle p^2 \right\rangle-2\left| m \right|(2l+d-2) \left\langle r^{-2} \right\rangle\\ &=\frac{4}{\eta^3} \left[ \eta-\left| m \right|\right], \hspace{6mm} d \geq 2.
\label{eqI_cap1:Icp}
\end{align}
for the position Fisher information of an arbitrary state $(n,l\equiv\mu_1,\mu_2,...,\mu_{d-1})$ of the system. Similar expressions can be obtained for the variance and Fisher information of the $d$-dimensional hydrogen atom  in momentum space \cite{Dehesa2010}. Indeed, one has
\begin{equation}\label{eqI_cap1:<p2><p4>}
\left\langle p^2 \right\rangle=\frac{1}{\eta^2}
\end{equation}
and 
\begin{align}\label{eqI_cap1:ImomT_2}
F\left[ \gamma \right]&=4 \left\langle r^2 \right\rangle-2\left| m \right| (2l+d-2) \left\langle p^{-2} \right\rangle\\ &=2 \eta^2 \left[5 \eta^2-3L(L+1) -\left| m \right|(8\eta-6L-3)+1\right]; \hspace{6mm} d \geq 2.
\end{align}
for the momentum expectation value $\left\langle p^2 \right\rangle$ and the momentum Fisher informations of an arbitrary state $(n,l,\left\lbrace \mu \right\rbrace)$, respectively. Surprisingly, the mean momentum expectation value $\left\langle p \right\rangle$ is not yet explicitly known for any state, but only for very-high lying (i.e., Rydberg) states (where the value is $\frac{2}{\pi \eta}$ \cite{Aptekarev2010}) and a few low-lying states of $n\textit{s}$ and circular types (see Appendix). Let us also mention that with the previous considerations about the variance, and since the Fisher informations of the $d$-dimensional probability density $\rho(\vec{r})$ and the one-dimensional density $\rho(r)$ are equal for unconfined spherically symmetrical systems, then one has that the Cr\'amer-Rao products $(\left\langle r^2 \right\rangle - \left\langle r\right\rangle^2) \times F[\rho]$ and $(\left\langle p^2 \right\rangle - \left\langle p\right\rangle^2) \times F[\gamma]$ are upper bounded by unity. Moreover, the modified Cr\'amer-Rao products $\left\langle r^2 \right\rangle \times F[\rho]\geq d^2$ and $\left\langle p^2 \right\rangle \times F[\gamma] \geq d^2$ are also fulfilled \cite{Dembo1991,Dehesa2007a,Dehesa2012}; in both cases the minimal bound is reached by the ground state of the d-dimensional harmonic oscillator.

Finally, from the general position expressions (\ref{eqI_cap1:varianza_rho}) and (\ref{eqI_cap1:Icp}) and the corresponding ones in momentum space, we can obtain the values gathered in Table I of the position and momentum variance and Fisher information for the 1s, 2s, 2p and 3d quantum states of the free 2D-HA (where $d=2$). Indeed, with $L=m+\frac{d-3}{2}$ and $d=2$ one has for example that 
\begin{equation}\label{posvarF}
V[\rho] = \frac{1}{8} (2 n^4- 4 n^3+7 n^2-5n-2 m^4+ m^2 +1),\quad F[\rho] = \frac{32(n-m-\frac{1}{2})}{(2n-1)^3}
\end{equation}
for the variance and Fisher information of any quantum state $(n, m)$ of the free 2D-HA in position space, respectively, and 
\begin{equation}
V\left[ \gamma \right] =   \frac{4}{(2n-1)^2}\left(1-\frac{\Gamma\left(n+\frac12\right)^4}{n^2\,\Gamma(n)^4}\right), \quad F\left[ \gamma \right] = \frac{1}{2} (2n-1)^2 (n+2),
\end{equation}
for the variance and Fisher information of any quantum state $(n, n-1)$ of the free 2D-HA in momentum space, respectively. See also Appendix, where you also have  $V\left[ \gamma \right] (2\textit{s})= \frac49-\frac{\pi^2}{64}\simeq  0.2902$. Moreover, taking into account Eq. (\ref{eq:cramer}), the last two columns of Table \ref{tabla:alfa1} collect the position and momentum Cr\'amer-Rao complexity measures for the four states of the free $2D$-HA under study.\\ 
\begin{table}[h!]
	\caption{Variance, Fisher information and Cr\'amer-Rao complexity for various low-lying states of the free 2D-HA in position and momentum spaces.}
	\centering
	\begin{tabular}{ c c c c c c c } 
		\hline
		State & $V[\rho]$ & $V[\gamma]$ & $F[\rho]$ & $F[\gamma]$ & $C_{CR}[\rho]$& $C_{CR}[\gamma]$\\ [0.5ex] 
		\hline\hline
		1s& 0.1250  & 1.5326  & 16.0000 &  1.5000  & 2.0000 & 2.2989  \\  
		2s& 2.3750  & 0.2902  & 1.7777  &  58.2000 & 4.2220 & 16.8896 \\
		2p& 2.2500  & 0.0975  & 0.5925  &  18.0000 & 1.3331 & 1.7550  \\
		3d& 9.3750  & 0.0245  & 0.1280  &  62.5000 & 1.2000 & 1.5312  \\[1ex]  
		\hline
	\end{tabular}
	\label{tabla:alfa1}
\end{table}

\section{Confined 2D-Hydrogen atom: Computational method.}
\label{sec:tres}
In this Section it is described the computational method used to compute the electronic wavefunctions and the associated probability densities for the stationary states $(n, m)$ of the confined two-dimensional hydrogen atom (i.e., an electron moving around the nucleus in a circular region of radius $r_{0}$  with impenetrable walls and the nucleus clamped at the center) in both position and momentum spaces. This confined 2D-HA obeys the Schr\"{o}dinger equation (\ref{schrodinger}) with $d=2$ and $\vec{r}  =  (r,\theta)$, where $r \equiv |\vec{r}|\in [0 , r_{0}]$ and $\theta\in [0 , 2\pi)$; so, their stationary states are characterized by two quantum numbers $(n, m)$ with $n=1,2,\dots$ and $m=0,1,\dots,n-1$. The states with $m=n-1$ are usually called by circular states. This equation cannot be solved in an analytical way, except for the limiting case  $r_{0}\rightarrow \infty$ which corresponds to the free (i.e., unconfined) system already considered in Section II for $d\, (d\geq2)$ dimensions.

\begin{figure}
	\centering
	\begin{minipage}{1\linewidth}
		\includegraphics[width=\linewidth]{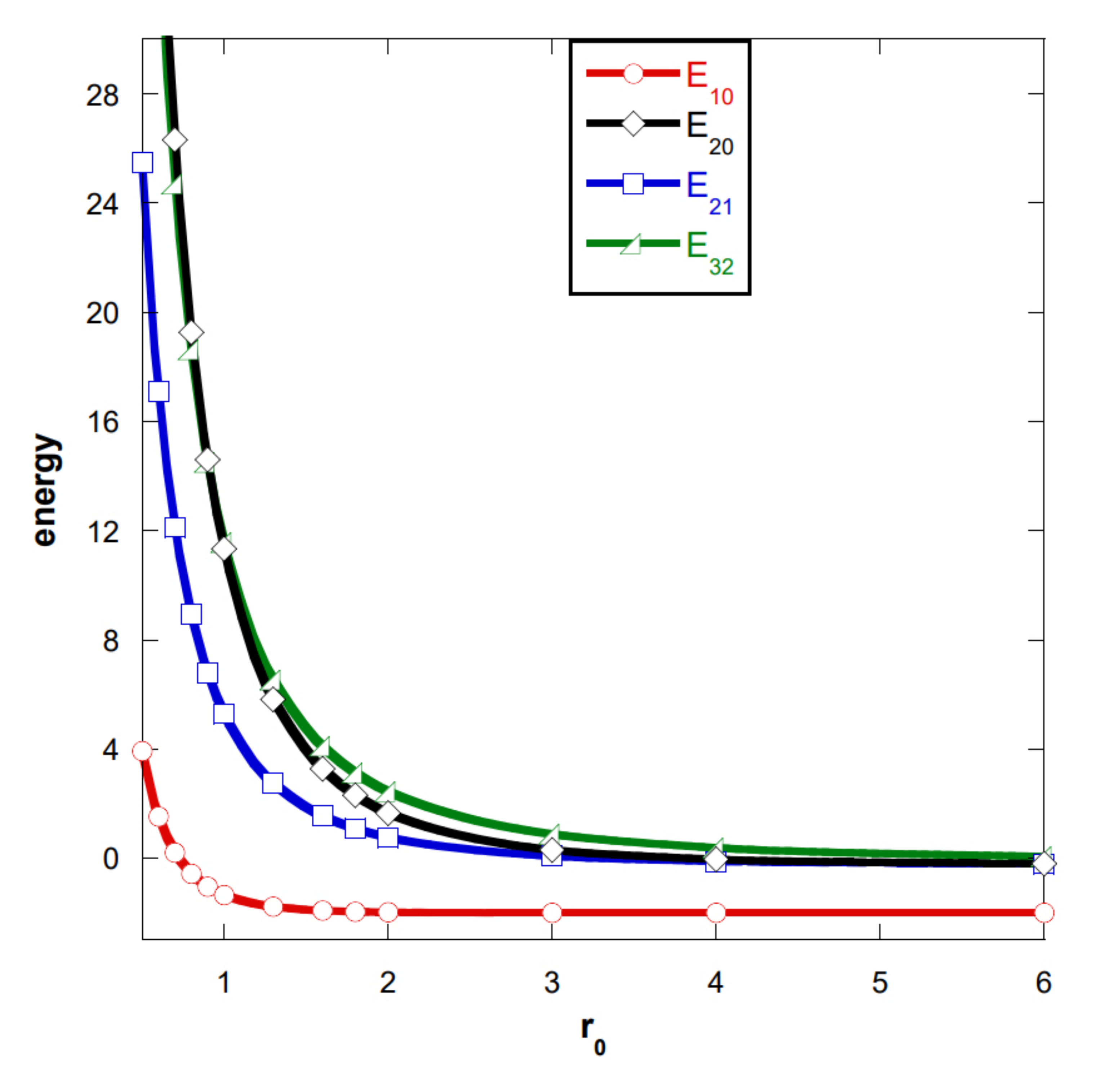}
	\end{minipage}
	\caption{Confinement dependence of the energies $E_{10}, E_{20}, E_{21}$ and $E_{32}$ for the 1s, 2s, 2p and 3d states of the 2D-HA, respectively. Atomic units have been used.}
	\label{Fig1}
\end{figure}

To compute the eigensolutions $\Psi_{n,m}^{(r_0)}(\vec r;\alpha)$ of the Schr\"{o}dinger equation of the confined 2D-HA, we have used the variational methodology described by Aquino et al \cite{Aquino1998} in the two-dimensional case and by Rojas et al \cite{Rojas2019} and by Marin and Cruz \cite{Marin}, Nascimento et al \cite{Nascimento2018}, Rojas et al \cite{Rojas2019} and Jiao et al \cite{Jiao2017} in the three-dimensional case. Then, we obtain
\begin{equation}\label{confined_wf}
\Psi_{n,m}^{(r_0)}(\vec r;\alpha)=R_{n,m}^{(r_0)}(r;\alpha) \,\frac{e^{im\theta}}{\sqrt{2\pi}},
\end{equation}
where $\alpha$ is a variational parameter, and the approximate radial part $R_{n,m}^{(r_0)}(r;\alpha)$ is given by  
\begin{equation}\label{confined_radial_wf}
R_{n,m}^{(r_0)}(r;\alpha)=N'_{nm}(\alpha)\,e^{-\alpha r} \left( \alpha r\right)^{\lvert m\rvert}L_{n-\lvert m\rvert-1}^{2\lvert m\rvert}\left(\alpha r\right)\,\chi^{(r_0)}(r)
\end{equation}
where the  cut-off function $\chi^{(r_0)}(r)=\left(1-\frac{r}{r_0}\right)$, 
$N'_{nm}(\alpha)$ denotes the normalization constant, and the optimized values of $\alpha$ are variationally derived. For further details about the role of the cut-off function we refer to some careful and systematic studies recently published \cite{Nascimento2018,Rojas2019}. The energies of the corresponding quantum states are numerically obtained as a function of the confinement radius $r_0$ by minimizing the functional $E(\alpha)=\langle\Psi|H|\Psi\rangle$ with respect to the variational parameter $\alpha$. The confinement dependence of the first few low-lying states $E_{10}, E_{20}, E_{21}$ and $E_{32}$ is shown in Figure \ref{Fig1}. 
We observe that for large values of $r_0$ these energies decrease monotonically towards to their respective values for the free case. The free constancy is reached for a state-dependent critical value $r_c$ of $r_0$, having the values $2$ a.u. (1s), 3 a.u. (2s,2p) and 5 a.u. (3d). Moreover, for all quantum states it happens that the greater the confinement (i.e., the smaller $r_0$), the greater the energy. Note, nevertheless, that the confinement-dependent energetic lines $E_{20}$ and $E_{32}$ cross each other at $r_{0}\simeq 1$ a.u., giving rise to the so-called (2s;3d) inversion. In fact, this is a common feature in systems confined by cavities with impenetrable walls.

The associated wavefunctions of the confined 2D-HA system in momentum space are determined by computing the Fourier transform of the position wavefunctions $\Psi_{n,m}^{(r_0)}(\vec r;\alpha)$ given by Eq. \eqref{confined_wf}, obtaining 
\begin{equation}
\begin{split}
\nonumber
\Phi^{(r_{0})}_{n,m}(\vec p)&=\frac{1}{2\pi}\int_{\mathbb R_2}\Psi^{(r_{0})}_{n,m}(r;\alpha)e^{-i\vec{p}\cdot\vec{r}} \text{d} \vec r\\
&=(2\pi)^{-\frac32}\int_{0}^{r_{0}} R^{(r_{0})}_{n,m}(r;\alpha)\, r\, \left[\int_{0}^{2\pi} e^{im\theta} e^{-i\vec{p}\cdot\vec{r}}\text{d}\theta\right]\,\text{d}r,
\end{split}
\end{equation}
which can be expressed as
\begin{equation}\label{momentum_wf}
\Phi^{(r_{0})}_{n,m}(\vec p)=\frac{i^{3m}\,e^{im\theta_p}}{(2\pi)^{1/2}}\int_{0}^{r_{0}}R^{(r_{0})}_{n,m}(r;\alpha)\,J_m(r\,p)\, r\text{d}r,
\end{equation}
where we have taken into account the integral representation of the Bessel function, $J_{n}(z)=\frac1{2\pi \,i^n}\int_0^{2\pi}e^{i\,n\tau}e^{i\,z\,cos\tau}\,d\tau$, and its parity property, $J_n(-z)=(-1)^nJ_n(z)$.\\

Finally, the position and momentum probability densities of the 2D-HA are given by
\begin{equation}\label{posden}
\rho_{n,m}(\vec r;r_0) = \left|\Psi_{n,m}^{(r_0)}(\vec r;\alpha)\right|^2; \quad \gamma_{n,m}(\vec p;r_0) = \left|\Phi_{n,m}^{(r_0)}(\vec p;\alpha)\right|^2,	
\end{equation}
respectively, which are the basic variables of the information theory of the two-dimensional confined hydrogenic system. They will be used to compute the dispersion- and complexity-like quantities of the system in the next sections.

\section{Variance and Cr\'amer-Rao complexity of 2D-Hydrogenic states}
\label{sec:cuatro}
In this section we study the confinement dependence of the variance, the Fisher information and the Cr\'amer-Rao complexity measure for the 1s, 2s, 2p and 3d states of the two-dimensional confined hydrogenic atom in both position and momentum spaces. These measures quantify the confinement effects on the concentration around the mean value and the gradient content of the charge and momentum delocalization of the system in an individual and joint manner, respectively. 

\subsection{The variance}

Since
\begin{equation}\nonumber
\left\langle r^\alpha \right\rangle \equiv \int_{\mathbb R_2} r^\alpha \rho_{n,m}(\vec r;r_0)\, d\vec{r}=\int^{\infty}_{0}r^{\alpha+d-1} |R^{(r_{0})}_{n,m}(r;\alpha)|^2 \, dr\\\label{eqI_cap1:<ra>},
\end{equation}
the variance of the probability density for a generic quantum state $(n, m)$ of the confined 2D-HA is given by 
\begin{equation}
V[\rho_{n,m}(\vec r;r_0)] = \int_0^{r_0} r^3\,|R^{(r_{0})}_{n,m}(r;\alpha)|^2\,\text{d}r - \left(\int_0^{r_0} r^2\, |R^{(r_{0})}_{n,m}(r;\alpha)|^2\,\text{d}r\right)^2
\end{equation}
in position space and by
\begin{equation}
V[\gamma_{n,m}(\vec r;r_0)] = \int_0^{r_0} p^3\,|\phi^{(r_{0})}_{nm}(p)|^2\,\text{d}p - \left(\int_0^{r_0}p^2\, |\phi^{(r_{0})}_{nm}(p)|^2\,\text{d}p\right)^2
\end{equation}
in momentum space, where the symbol $\phi^{(r_{0})}_{nm}(p)$ denotes the radial part of the momentum eigenfunction.

\begin{figure}\centering
	\begin{minipage}{1\linewidth}
		\includegraphics[width=\linewidth]{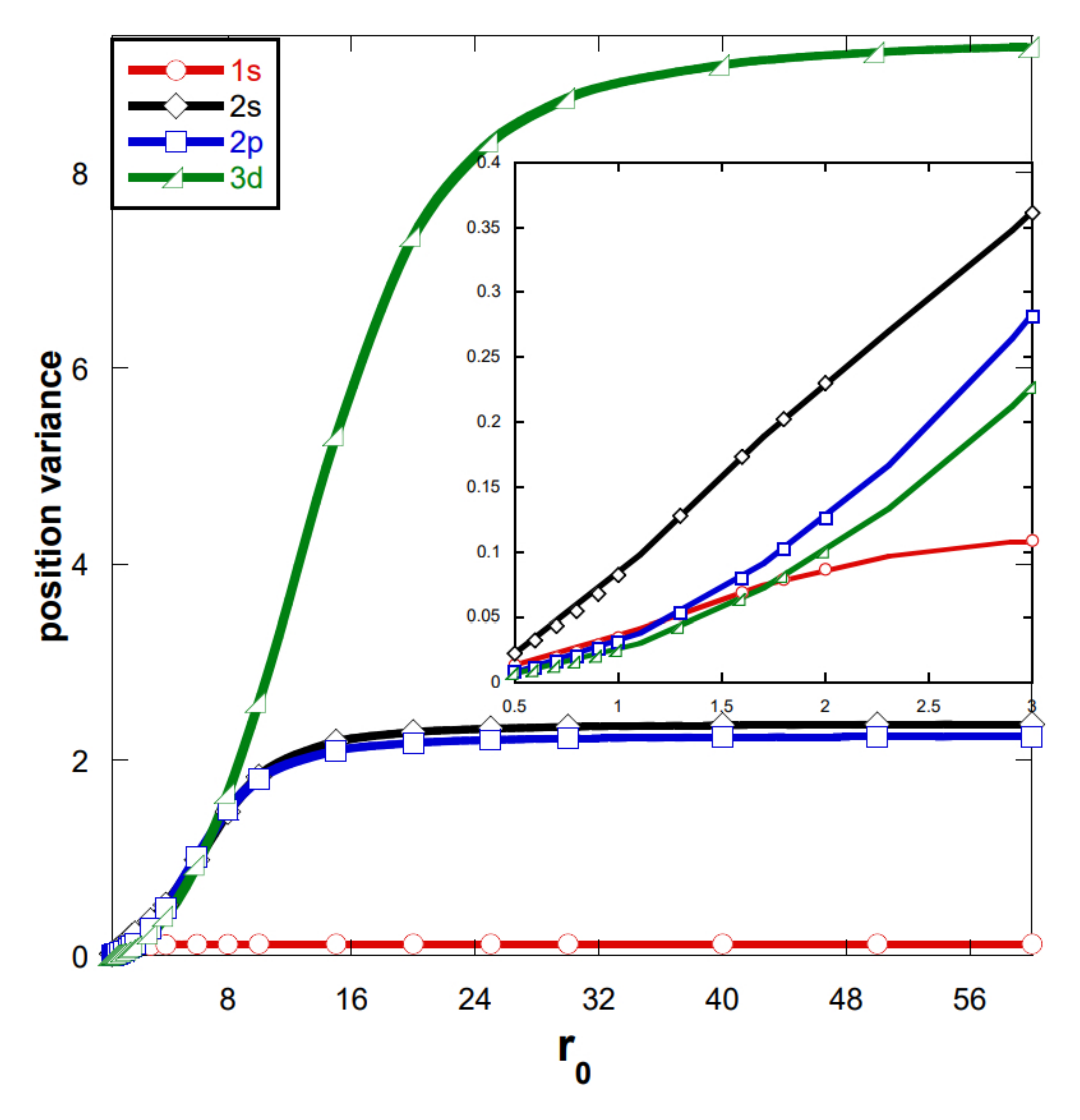}
	\end{minipage}
	\caption{Confinement dependence of the variance for the 1s, 2s, 2p and 3d states of the confined 2D-HA in position space. Atomic units have been used.}
	\label{FIG2}
\end{figure}

\begin{figure}\centering
	\begin{minipage}{1\linewidth}
		\includegraphics[width=\linewidth]{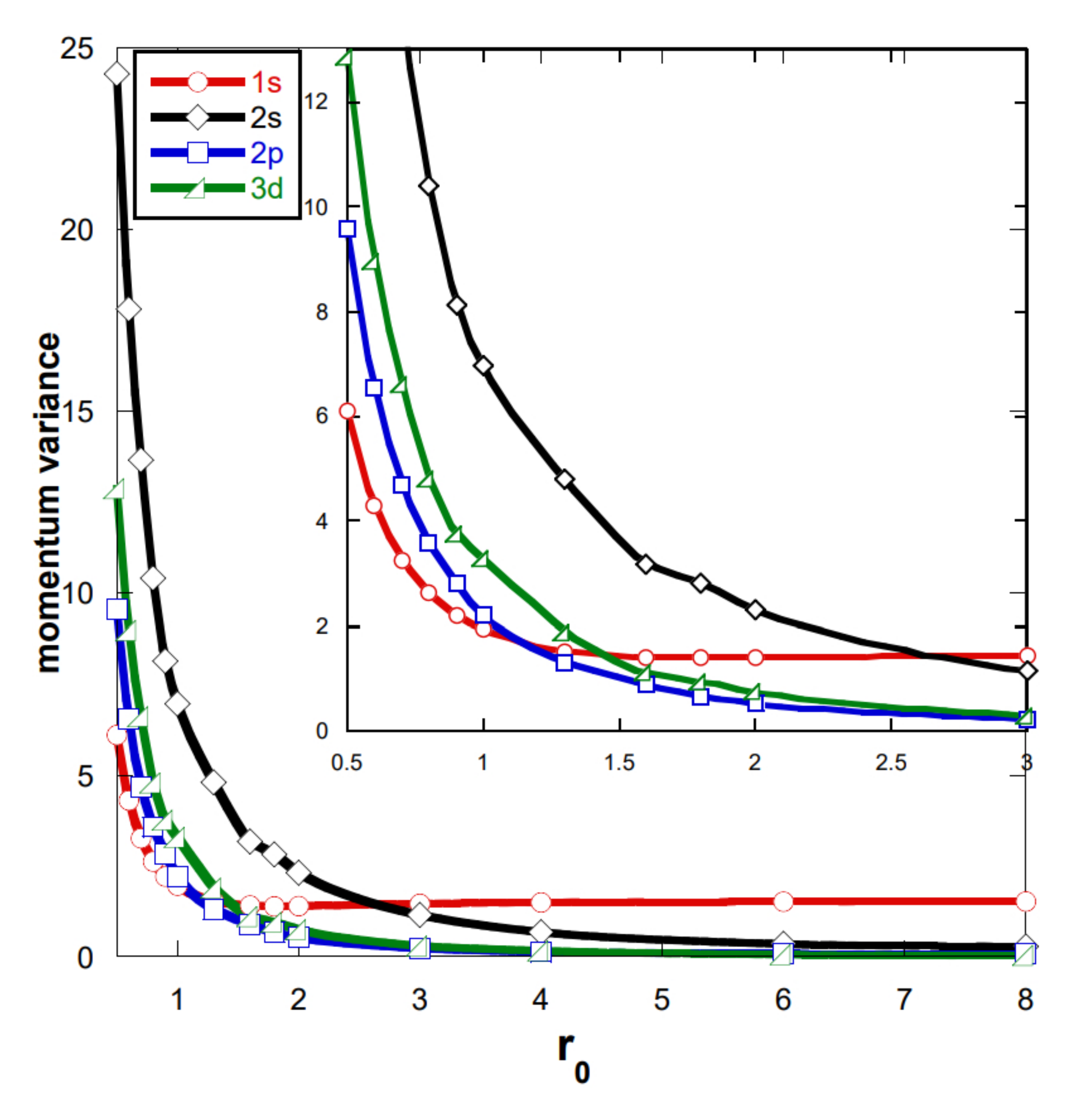}
	\end{minipage}
	\caption{Confinement dependence of the variance for the 1s, 2s, 2p and 3d states of the 2D-HA in momentum space. Atomic units have been used.}
	\label{Fig3}
\end{figure}

In Figures \ref{FIG2} and \ref{Fig3} we show the variance for the quantum states 1\textit{s}, 2\textit{s}, 2\textit{p} and 3\textit{d} of the confined 2D-HA as a function of the confinement radius $r_0$ in both position and momentum spaces, respectively. First we observe that both position  and momentum values for the variance increase and decrease, respectively, to the rigorously calculated (as shown above in Table I) free values when the confinement radius $r_0$ increases, reaching the free ones at a state-dependent critical value $r_c$ around $8$ \,a.u.$(1\textit{s})$, $20$\, a.u.$(2\textit{p})$, $30$\, a.u.$(3\textit{d})$ and $60$\, a.u.$(2\textit{s})$ in position space and around $2$\,a.u.$(1s)$, $3$\, a.u. $(2p)$,  $5$\, a.u.$ (3d)$ and $7$\, a.u. $(2s)$ in momentum space. \\ 

Note that in the (strong) confinement region $(0,r_c)$ the position variance has an increasing first-convex-then-concave behavior, while the momentum variance go down monotonically to their corresponding free constant values in a concave way. Moreover, from Figure \ref{FIG2}, one also observes that for the circular states (1s, 2p, 3d) the position variances of the electron distribution not only decrease when the confinement radius decreases from the critical values just mentioned, but also they cross each other; for example, the variance of the ground state has two crossings, one with the variance of the state 2p and another one with the variance of the state 3d. In addition, the variance of the state 3d has a crossing with the variance of each of the remaining states. And from Figure \ref{Fig3} we realize  that for all quantum states the greater the confinement (i.e., the smaller $r_0$), the greater the momentum variance. This is a behavior similar to the confinement dependence of the energy shown in Figure \ref{Fig1}. Note, however, that confinement effect on the momentum variance provokes three crossings at around $1.2$ a.u., $1.8$ a.u. and $2.7$ a.u. of (1s;2p), (1s;3d) and (1s;2s) character, respectively, while in Figure \ref{Fig1} we only had an energy (3d;2s) inversion at around $1$ a.u.

\subsection{The Fisher information}

The Fisher information for the stationary states of the two-dimensional confined hydrogenic atom, which are characterized by the position and momentum probability densities $\rho(\vec r;r_0)$ and $\gamma(\vec p;r_0)$ respectively, defined by Eq. \eqref{posden}, is given \cite{Fisher1925} (see also \cite{Dehesa2011}) by
\begin{equation}
F_\rho(r_0)=\int_{\mathbb R_2}\frac{\left|\nabla\rho(\vec r;r_0)\right|^2}{\rho(\vec r;r_0)}\,d\vec r; \quad  F_\gamma(r_0)=\int_{\mathbb R_2}\frac{\left|\nabla\gamma(\vec p;r_0)\right|^2}{\gamma(\vec p;r_0)}\,d\vec p,
\end{equation}
which satisfy the uncertainty relation $F_\rho\times F_\gamma\ge 4\times 2^2 = 16$ for the real wavefunctions of the system \cite{Sanchez-Moreno2011}, what in our case occurs for the $1s$ and $2s$ states only. The Fisher information $F_\rho(r_0)$ for the state $(n,m)$ of a 2D-HA is a local measure of spreading of the density $\rho_{n,m}(\vec r;r_0)$ because it is a gradient functional of $\rho_{n,m}(\vec r;r_0)$. The higher this quantity is, the more localized is the density, the smaller is the uncertainty and the higher is the accuracy in estimating the localization  of the particle. Recently (see Figure 3 of \cite{Estanon2020}) the Fisher information of the 2D-HA has been computed for the 1s, 2s, 2p and 3d states. Therein, we found that the Fisher information decreases (position) and increases (momentum) when $r_0$ is increasing, so that they tend broadly and fastly to the free values (analytically calculated in section II and numerically given in Table I) in such a way that the Fisher-information-based uncertainty relation for the $1s$ and $2s$ states is always fulfilled because they are described by real wavefunctions.

\subsection{The Cr\'amer-Rao complexity}

The Cr\'amer-Rao complexity measure for the stationary states of the 2D-HA, according to Eqs.(\ref{eq:cramer}) and (\ref{posden}), are given by 
\begin{equation} \label{eq:crtwo}
C_{\text{CR}}[\rho_{n,m}(\vec r;r_0)] =F[\rho_{n,m}(\vec r;r_0)]\times V[\rho_{n,m}(\vec r;r_0)]
\end{equation}
and
\begin{equation}
C_{\text{CR}}[\gamma_{n,m}(\vec p;r_0)] =F[\gamma_{n,m}(\vec p;r_0)]\times V[\gamma_{n,m}(\vec p;r_0)]	
\end{equation}
in both position and momentum spaces, respectively. These measures quantify the combined balance of the gradient content of the density jointly with its concentration around the centroid in both conjugated spaces. So, they are statistical complexities of local-global character.\\

\begin{figure}\centering
	\begin{minipage}{1\linewidth}
		\includegraphics[width=\linewidth]{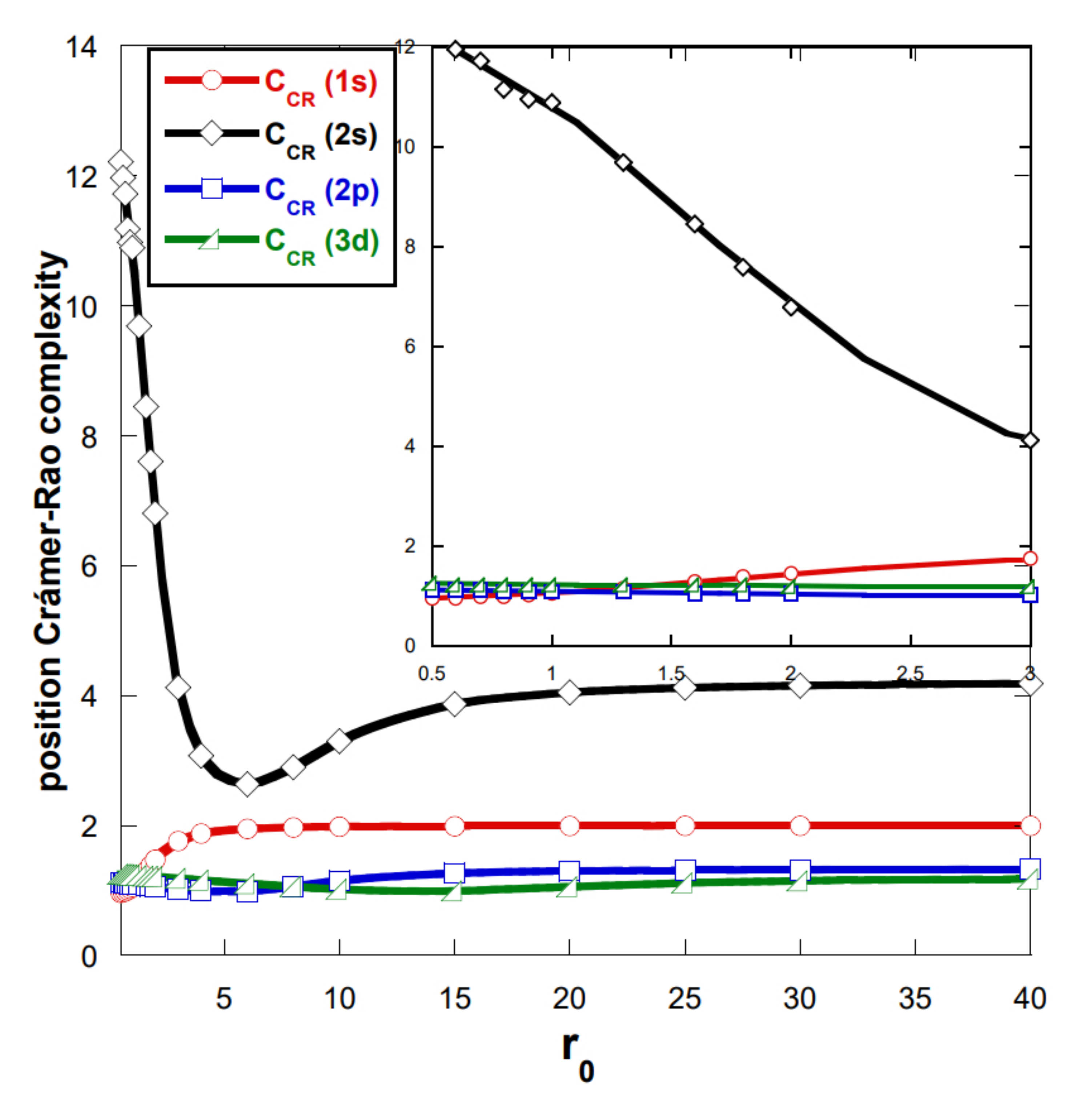}
	\end{minipage}
	\caption{Confinement dependence of the Cr\'amer-Rao complexity measure for the 1s, 2s, 2p and 3d states of the confined 2D-HA in position space. Atomic units have been used.}
	\label{Fig4}
\end{figure}

\begin{figure}\centering
	\begin{minipage}{1\linewidth}
		\includegraphics[width=\linewidth]{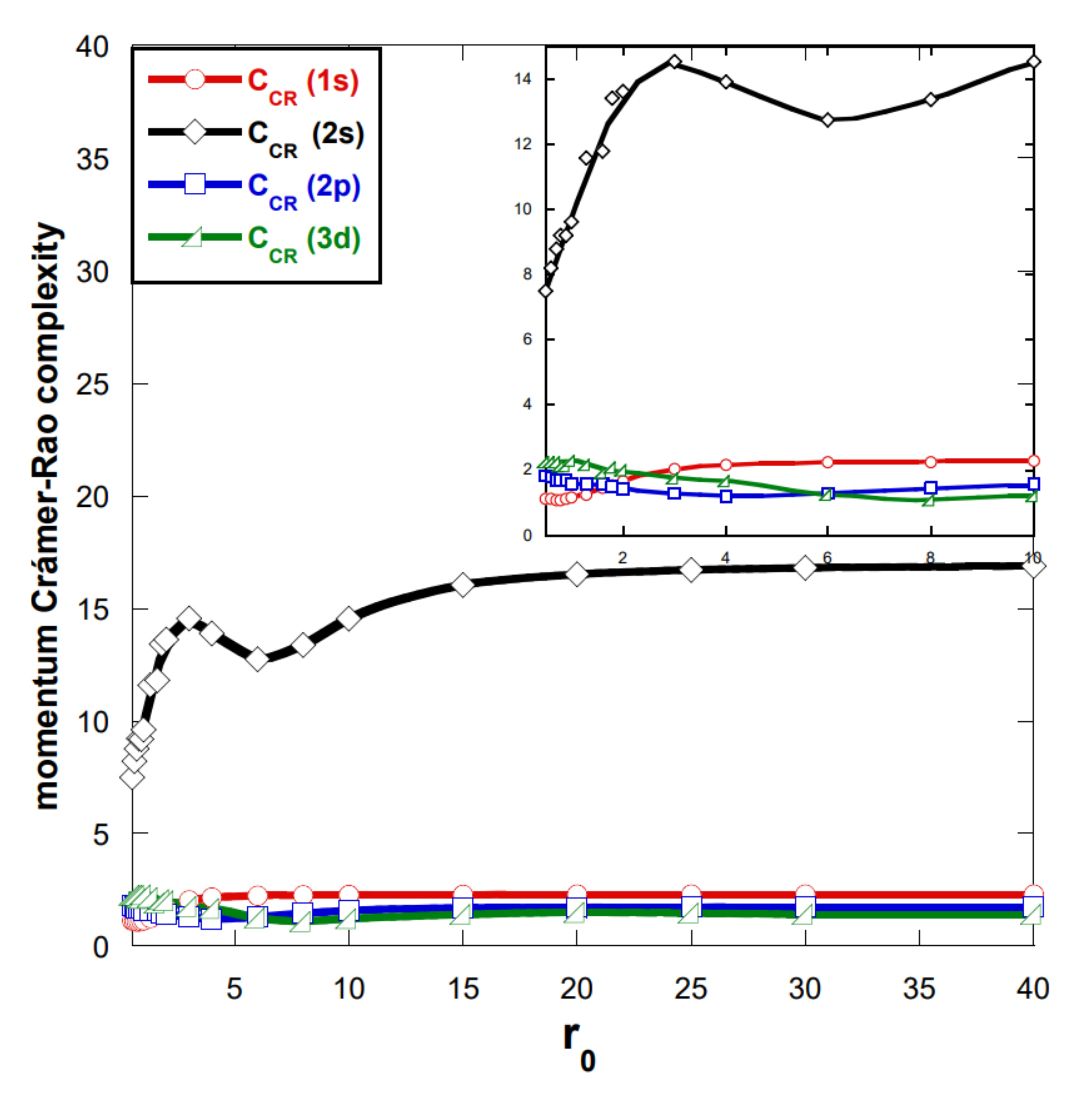}
	\end{minipage}
	\caption{Confinement dependence of the Cr\'amer-Rao complexity measure for the 1s, 2s, 2p and 3d states of the 2D-CHA in momentum space. Atomic units have been used.}
	\label{Fig5}
\end{figure}

In Figures \ref{Fig4} and \ref{Fig5} we show the values of these Cr\'amer-Rao measures for the quantum states 1\textit{s}, 2\textit{s}, 2\textit{p} and 3\textit{d} of the confined 2D-HA as a function of the confinement radius $r_0$ in both position and momentum spaces, respectively. First we observe that confinement does distinguish charge disorder and momentum complexity for the four quantum states under study. Moreover, these position and momentum measures appear to be lowerbounded by unity for all values of the confinement radius; that is, not only asymptotically where the system becomes unconfined and then this property is rigorously fulfilled as already mentioned at the end of section II. 

Note in Figure \ref{Fig4} that the position Cr\'amer-Rao measures behave differently for the four states when $r_0$ increases, although all of them tend towards constancy when $r_0$ is sufficiently large; that is, when confinement is sufficiently weak and then the 2D-HA system becomes practically free. Such a constancy is reached at $15$ a.u. (or even earlier) for the circular states and at $25$ a.u. for the state $2s$. Above this critical confinement radius the Cr\'amer-Rao measure satisfies the inequality chain 
$$C_{CR}[3d] < C_{CR}[2p] < C_{CR}[1s] < C_{CR}[2s].$$ Note that these constant free values for the position Cr\'amer-Rao measure are 1.2000 (3d), 1.3331 (2p), 2.0000 (1s) and 4.2220 (2s), which coincide with the values collected in the last two columns of Table \ref{tabla:alfa1} which were obtained in a rigorous analytical way; this is a further checking of our numerical results.\\
For stronger confinements the situation gets much more involved. Note that the Cr\'amer-Rao ordering for the circular states gets altered for values of $r_0$ less than $10$ a.u., and the Cr\'amer-Rao measure $C_{CR}[2s]$ shows a minimum at around $6$ a.u. because of the delicate balance of the charge concentration around the centroid and the oscillatory character of the electron density at such a value of the confinement radius.\\

Similar considerations can be done from Figure \ref{Fig5} for the momentum Cr\'amer-Rao complexity measures of the states 1\textit{s}, 2\textit{s}, 2\textit{p} and 3\textit{d}. Here again we observe that for all ground and excited states this quantity tends towards the corresponding constant known free values when $r_0$ increases (i.e., when the confinement is weaker and weaker), so that it is fulfilled the same complexity ordering pointed out previously in position space. This constancy is reached at $15\, a.u.$ (or even earlier) for circular states and at $25$ a.u. for the state $2s$, as in position space. This behavior towards free constancy is, of course, different for each quantum state. In the ground state the measure smoothly increases when $r_0$ increases up until the free constant value which is reached at $4\,a.u.$. In the first excited state ($2s$) the momentum Cr\'amer-Rao complexity has a completely different behavior with respect to the position one when $r_0$ increases: the former measure has a pronounced maximum at about $r_0 = 5\, a.u.$, from which it oscillates to the free constancy. For the circular states $2p$ and $3d$ the values of this momentum measure smoothly vary towards the corresponding free constant values rigorously calculated in Section \ref{sec:freeHA} and collected in Table I; remark that they cross each other at $r_0 = 6\, a.u.$ All this suggests a strong dependence of the Crámer-Rao complexity measure on the quantum-number difference $n-|m|.$\\

For stronger confinements (say, when $r_0 < 10 a.u.$) the previous complexity ordering changes, mainly among the momentum complexities of the ground state and the other two circular states; this is basically because of the delicate interplay of the momentum concentration around the centroid and the gradient content of the electron density.

\section{Conclusions}

In this work the internal disorder of the confined two-dimensional hydrogen atom is studied in a few low-lying quantum states (1\textit{s}, 2\textit{s}, 2\textit{p}, 3\textit{d}) by means of the variance and the Cr\'amer-Rao complexity measure in both position and momentum spaces as a function of the confinement radius $r_0$. These measures, which do not depend on the energy but on the eigenfunction of the state, quantify the pointwise concentration of the electronic charge around the centroid and the combined balance of this concentration and the gradient content of the electron density all over the confinement region, respectively. \\
We have found that the position and momentum variances increase and decrease, respectively, when the confinement becomes weaker and weaker (that is, when $r_0$ is increasing), so that for a critical confinement radius onwards the variances have the values of the free (unconfined) 2D-HA which are rigorous and analytically determined. This critical radius is bigger in position space than in momentum space for each state. Moreover, the variances go monotonically up (in position space) and down (momentum space) without crossing, except for the ground state, to the corresponding free values. Then, the greater the confinement (i.e.,the smaller $r_0$), the smaller the position variance and the greater the momentum variance.\\

We have also shown here that confinement does distinguish complexity of the 2D-CHA for all stationary states by means of the Cr\'amer-Rao measure in the two conjugated spaces. These quantities tend in various ways towards the corresponding constant free values for both ground and excited states when the confinement radius $r_0$ increases. This constancy, which is also analytically calculated, is reached at a critical confinement radius, which is much less for circular states than for the state 2s basically because of the bigger relative gradient content. So, this complexity measure best detects the confinement effects when the impenetrable wall of the system is located at a gradually smaller critical radius.

The present results together with some recent efforts on information entropies of global (Shannon, R\'enyi, Tsallis) \cite{Mukherjee2018,Mukherjee2018a,Aquino2013} and local (Fisher, relative Fisher) \cite{Mukherjee2018b,Mukherjee2018c,Aquino2013,Wu2020,Yu2017} character for numerous excited states of \textit{three}-dimensional free and confined hydrogen-like systems, illustrate how and how much confinement is crucial not only for the energy spectrum of multidimensional hydrogen \cite{Ping2019,Rojas2018} but also for its eigenfunction-dependent information-theoretical properties which control all the chemical and physical properties. 


\section*{Funding information}
This work has been partially supported by the grant FIS2017-89349P of the Agencia Estatal de Investigaci\'on (Spain) and the European Regional Development Fund (FEDER). \\

\section*{Author contributions}

N.A., C.R.E. and J.S.D. contributed in conceptualization, data curation, investigation, and writing of the original draft of the manuscript. C.R.E. and D.P.C. performed numerical analysis, and edited the manuscript.

\appendix

\section{Dispersion and Fisher information values of $n\textit{s}$ and circular states of the free $d$- and 2-dimensional hydrogenic atom}

The mean expectation value $\left\langle p \right\rangle$ of the momentum probability density of the free $d$-dimensional hydrogenic atom is not yet explicitly known for a generic stationary state $(n,l,\left\lbrace \mu \right\rbrace)\equiv(n,l\equiv\mu_1,\mu_2,...,\mu_{d-1})$ despite the efforts of various authors \cite{Assche2000,Adkins2019,Delbourgo2009}. However, its value can be analytically found for some specific states such as the $(n\textit{s})$ and circular states which are the \textit{leitmotiv} of this work. This is the purpose of this appendix. In addition we illustrate here the analytical computation of the other dispersion and Fisher information measures which we are interesting in. This is possible because the general expressions (\ref{eqI_cap1:Rnl}), (\ref{eqI_cap1:DenPos}), (\ref{eqI_cap1:Mnl}) and (\ref{eqI_cap1:DenMom}) strongly simplifies for such states as it is illustrated for convenience in the following. 

\subsection{n\textit{s} states}
For the (n\textit{s})-states (i.e., when $\mu_i=0$, $\forall \hspace{2mm} i=1,\dots,d-1$ ) one finds, from Eqs. (\ref{eqI_cap1:Rnl}), (\ref{eqI_cap1:DenPos}), (\ref{eqI_cap1:Mnl}) and (\ref{eqI_cap1:DenMom}), the expressions
\begin{equation}\label{eqI_cap1:rho_ns}
\rho(\vec{r}) (n\textit{s})=\frac{2^{2d-1}  \Gamma\left( \frac{d}{2}\right)}{\pi^{\frac{d}{2}} (2n+d-3)^{d+1}}  e^{-\frac{r}{\lambda}} \left| {\tilde{\cal{L}}}^{(d-2)}_{n-1} \left( \frac{r}{\lambda} \right) \right|^2,
\end{equation}
\begin{equation}\label{eqI_cap1:gamma_ns}
\gamma(\vec{p}) (ns)=\frac{(2n+d-3)^{d} \Gamma \left(\frac{d}{2} \right)}{8\pi^{\frac{d}{2}}} \left[  {\tilde{C}}^{\frac{d-1}{2}}_{n-1}\left( \frac{1-\eta^2 p^2}{1+\eta^2 p^2}\right) \right]^2,
\end{equation}
for the position and momentum probability densities of the free $d$-dimensional hydrogenic system, respectively. And, moreover, when $n=1$ we have the expressions
\begin{equation}\label{eqI_cap1:rho_gs}
\rho(\vec{r}) (1\textit{s})= \left(\frac{2}{d-1} \right)^d \frac{1 }{\pi^{\frac{d-1}{2}} \Gamma\left( \frac{d+1}{2} \right) } e^{-\frac{4}{d-1}r},
\end{equation}
\begin{equation}\label{eqI_cap1:gamma_gs}
\gamma(\vec{p}) (1\textit{s})= \frac{(d-1)^d \Gamma \left(\frac{d+1}{2} \right)}{\pi^{\frac{d+1}{2}}} \frac{1}{\left( 1+\frac{(d-1)^2}{4} p^2 \right)^{d+1}},
\end{equation}
for the position and momentum probability densities of the ground state of the $d$-dimensional free hydrogen atom. From here we find that
\begin{equation}\label{eqI_cap1:<pa>_gs}
\left\langle p^\alpha \right\rangle (1\textit{s})=\left( \frac{2}{d-1} \right)^\alpha \frac{2 \Gamma\left(\frac{d-\alpha}{2}+1 \right) \Gamma\left(\frac{d+\alpha}{2} \right) }{d\Gamma^2\left(\frac{d}{2} \right)}; \hspace{5mm}-d<\alpha<d+2
\end{equation}
so that
\begin{equation}\label{eqI_cap1:<p><p2>_gs}
\left\langle p \right\rangle (1\textit{s})=\frac{4}{d(d-1)}  \frac{\Gamma^2 \left( \frac{d+1}{2}\right) }{\Gamma^2 \left( \frac{d}{2}\right)}, \hspace{1cm} \left\langle p^2 \right\rangle (1\textit{s})=\left( \frac{2}{d-1}\right)^2,
\end{equation}
and the variance becomes
\begin{equation}\label{eqI_cap1:varianza_gamma_gs}
V\left[ \gamma \right] (1\textit{s})=\left( \frac{2}{d-1}\right)^2 \left[ 1-\frac{4}{d^2}\left( \frac{\Gamma \left( \frac{d+1}{2}\right) }{\Gamma \left( \frac{d}{2}\right)} \right)^4\right].
\end{equation}
Then, for $d=2$ one has $\left\langle p \right\rangle (1\textit{s})= \frac{\pi}{2}$ and $\left\langle p^2 \right\rangle (1\textit{s})= 4$ so that $V\left[ \gamma \right] (1\textit{s})= 4 - \frac{\pi^2}{4} = 1.5326$, which is given in Table I.
Similarly, we can calculate the momentum variance for states $(n\textit{s})$ of the two-dimensional hydrogen atom other than $n=1$, obtaining 
\begin{equation}\label{eqI_cap1:<p><p2>_gs}
\left\langle p \right\rangle (n\textit{s})= \sum_{j=0}^{n-1}(-1)^j\frac{2j+1}{2j+2} \frac{\left[\Gamma\left(j+\frac12\right)\right]^2}{\left[\Gamma\left(j+1\right)\right]^4}\frac{\Gamma(n+j)}{\Gamma(n-j)} ; \hspace{1cm} \left\langle p^2 \right\rangle (n\textit{s})=\left( \frac{2}{2n-1}\right)^2,
\end{equation}
and the variance straighforwardly follows from $V\left[ \gamma \right] (n\textit{s})=\left\langle p^2 \right\rangle (n\textit{s}) - \left\langle p \right\rangle^2 (n\textit{s})$. In particular, for $n=2$ one has that  $\left\langle p \right\rangle (2\textit{s})= \frac \pi8 \simeq 0.3927$ and $\left\langle p^2 \right\rangle (2\textit{s})=\frac{4}{9}$, so that the variance $V\left[ \gamma \right] (2\textit{s})= \frac49-\frac{\pi^2}{64} \simeq 0.2902$, which is given in Table I.  As well we can obtain the values
\begin{equation}\label{eqI_cap1:I_n0_00}
F\left[ \rho \right] (ns)=\left( \frac{2}{\eta}\right)^2; \hspace{7mm} F\left[ \gamma \right] (ns)=\eta^2\left[ 10 \eta^2-\frac{3}{2} (d-3)(d-1)+2 \right],
\end{equation} 
for the position and momentum Fisher information for the (n\textit{s})-states of the free $d$-dimensional hydrogen atom, respectively, so that we have for the ground state the values
\begin{equation}\label{eqI_cap1:I_n0_00}
F\left[ \rho \right] (1\textit{s})=\left( \frac{4}{d-1}\right)^2, \hspace{7mm}F\left[ \gamma \right] (1\textit{s})= \frac{1}{4}d(d+1)(d-1)^2
\end{equation}
Then, for $d=2$ the last few expressions give the values of the position and momentum  Fisher information for the 1s and 2s states of the free 2D-HA gathered in Table I. 

\subsection{Circular states}
Other particular states interesting for our purposes are the circular states ($n,\mu_1=\mu_2=\dots=\mu_{d-1}=n-1$) which, according to Eqs. (\ref{eqI_cap1:Rnl}), (\ref{eqI_cap1:DenPos}), (\ref{eqI_cap1:Mnl}) and (\ref{eqI_cap1:DenMom}), have the following probability densities
\begin{equation}\label{eqI_cap1:DenPos_circular}
\rho_{\mbox{circ}}(\vec{r}) =\frac{2^{d+2-2n}}{\pi^{\frac{d-1}{2}}(2n+d-3)^{d} \Gamma(n) \Gamma \left(n+\frac{d-1}{2} \right)  } \times e^{-\frac{r}{\lambda} } \left( \frac{r}{\lambda} \right)^{2n-2} \prod^{d-2}_{j=1} \left( \sin \theta_j \right)^{2n-2},
\end{equation}
and 
\begin{equation}\label{eqI_cap1:DenMom_circular}
\gamma_{\mbox{circ}}(\vec{p})= \frac{2^{2n-2} (2n+d-3)^d \Gamma \left( n+\frac{d-1}{2}\right)}{\pi^{\frac{d+1}{2}} \Gamma(n)} \frac{(\eta p)^{2n-2}}{(1+\eta^2 p^2)^{2n+d-1}} \prod^{d-2}_{j=1} \left( \sin \theta_j \right)^{2n-2},
\end{equation}
for the position and momentum probability densities, respectively. \\
From these two expressions we can obtain all the dispersion, entropy-like and complexity-like quantities of the circular states of the $d$-dimensional free hydrogen atom as defined in section I. 
In particular we have the values
\begin{equation}\label{eqI_cap1:<ra>_circular}
\left\langle r^\alpha \right\rangle= \left( \frac{2n+d-3}{4} \right)^\alpha \frac{\Gamma(2n+d-2+\alpha)}{\Gamma(2n+d-2)}; \hspace{6mm} \alpha > -2n-d+2,
\end{equation}
\begin{equation}\label{eqI_cap1:<pa>_circular}
\left\langle p^\alpha \right\rangle= \left( \frac{2}{2n+d-3} \right)^\alpha \frac{ \Gamma \left( n+\frac{d+\alpha-2}{2} \right) \Gamma \left( n+\frac{d-\alpha}{2} \right) }{\left( n+\frac{d-2}{2}\right) \Gamma^2 \left( n+\frac{d-2}{2} \right) },\quad  -2n-d+2 < \alpha < 2n+d
\end{equation}
for the position and momentum expectation values of order $\alpha$, respectively. The use of these expressions with $\alpha=1$ and $2$ allows one to find the values for the position and momentum variance for the circular states, obtaining for example the value
\begin{equation}\label{eqI_cap1:varianza_gamma_gs}
V\left[ \gamma_{\mbox{circ}} \right](1s)=\left( \frac{2}{d-1}\right)^2 \left[ 1-\frac{4}{d^2}\left( \frac{\Gamma \left( \frac{d+1}{2}\right) }{\Gamma \left( \frac{d}{2}\right)} \right)^4\right].
\end{equation}
for the momentum variance of the ground state (1\textit{s}) of the free $d$-dimensional hydrogenic atom. Moreover, we find that
\begin{equation}
\langle p\rangle(d=2) =\frac{2 \Gamma^2(\frac{1}{2}+n)}{n (2 n-1) \Gamma^2(n)}            ; \quad \langle p^2\rangle (d=2) =   \frac{4 }{ (2 n-1)^2 }
\end{equation} 
so that the momentum variance for any circular state $(n, n-1)$ of a 2D-HA has the value
\begin{equation}
V\left[ \gamma_{\mbox{circ}} \right](d=2) =   \frac{4}{(2n-1)^2}\left(1-\frac{\Gamma\left(n+\frac12\right)^4}{n^2\,\Gamma(n)^4}\right).   
\end{equation}
The last three expressions for $d=2$ give the value $V\left[ \gamma \right](1s; d=2)= 4(1-\pi^2/16) = 1.5326$ as well as the momentum variances for the other circular states, 2p and 3d, of Table I. The rest of values for the position and momentum variances gathered in the second and third column of this table for the two-dimensional free hydrogen can be obtained in a similar manner.
For example, we have the values
\begin{equation}\label{eqI_cap1:I_rho_circular}
F\left[ \rho_{\mbox{circ}} \right] = \frac{2 (d-1)}{\eta^3}= \frac{16 (d-1)}{(2n+d-3)^3},
\end{equation}
\begin{equation}\label{eqI_cap1:I_gamma_circular}
F\left[ \gamma_{\mbox{circ}} \right] = \frac{1}{4}(2n+d-3)^2 \left[(2n+d)(d-1)+2 \right],
\end{equation}
for the position and momentum Fisher information, which for $d=2$ give the values 
\begin{equation}
F\left[ \rho_{\mbox{circ}} \right] = \frac{16}{(2n-1)^3}; \hspace{7mm} F\left[ \gamma_{\mbox{circ}} \right] = \frac{1}{2} (2n-1)^2 (n+2),
\end{equation}
for the position and momentum Fisher information of the free 2D-HA. For $n=1,2,,3$ they supply the corresponding values of the Fisher information of the circular states gathered in Table I; that is, $1\textit{s}, 2\textit{p}$ and $3\textit{d}$.\\
Finally, for completeness, since 
\begin{equation}\label{eqI_cap1:<ra>_circular}
\left\langle r^2 \right\rangle= \left( \frac{2n+d-3}{4} \right)^2 (2n+d-2) (2n+d-1); \quad \left\langle p^2 \right\rangle= \left( \frac{2}{2n+d-3} \right)^2 
\end{equation}
we also observe that
\begin{equation}\label{eqI_cap1:<r2>I_rho_circular}
\left\langle r^2 \right\rangle F\left[ \rho \right]= \frac{ (d-1) (2n+d-1)(2n+d-2)}{2n+d-3};\hspace{7mm} \left\langle p^2 \right\rangle F\left[ \gamma \right]= (2n+d)(d-1)+2,
\end{equation}
for the Cr\'amer-Rao uncertainty products for the circular states $(n,n-1)$ of the 2D-HA, which are always greater than $d^2$ as they should (see at the end of section II) \cite{Dembo1991,Dehesa2007a,Dehesa2012}.


\begin{thebibliography}{99}
	
	\bibitem{Sen2011} \textit{Statistical Complexities: Application to Electronic Structure} (Ed: K.D. Sen), Springer, Berlin \textbf{2011}.
	\bibitem{Herschbach1993}  \textit{Dimensional Scaling in Chemical Physics} (Eds: D.R. Herschbach, J. Avery, O. Goscinski, Kluwer, Dordrecht \textbf{1993}.
	
	\bibitem{Dehesa2011} J. S. Dehesa, S. López-Rosa,  D. Manzano, Entropy and complexity analyses of $D$-dimensional quantum systems. In \textit{Statistical Complexities: Application to Electronic Structure} (Ed: K.D. Sen), Springer, Berlin/Heidelberg, \textbf{2011}.
	
	\bibitem{Angulo2011} J.C. Angulo, J. Antol\'in, R.O. Esquivel, Atomic and Molecular Complexities: Their Physical and Chemical Interpretations. In \textit{Statistical Complexities: Application to Electronic Structure} (Ed: K.D. Sen), Springer, Berlin/Heidelberg, \textbf{2011}.
	
	\bibitem{Dehesa2010} J.S. Dehesa, S. L\'opez-Rosa, A. Mart\'inez-Finkelshtein, R.J. Yáñez, Int. J. Quantum Chem.  \textbf{2010}, 110, 1529.
	\bibitem{Fisher1925} R.A. Fisher, Proc. Cambridge Philos. Soc. \textbf{1925}, 22, 700.
	\bibitem{Shannon1948} C. E. Shannon, Bell Syst. Tech. J. \textbf{1948}, 27, 379.
	\bibitem{Gonzalez2003} R. González-Férez, J. S. Dehesa, Physical Review Letters \textbf{2003}, 91, 113001.
	
	\bibitem{Gonzalez2005} R. González-Férez, J. S. Dehesa, European Physical Journal D \textbf{2005}, 32, 39.
	\bibitem{Esquivel2015} R.O. Esquivel, M. Molina-Espiritu, S. Lopez-Rosa, C. Soriano-Correa, C. Barrientos-Salcedo, M. Kohout, J.S. Dehesa, ChemPhysChem \textbf{2015}, 16, 2571
	\bibitem{Esquivel2016} R.O. Esquivel, S. López-Rosa, M. Molina-Espíritu, J. C.Angulo, J.S. Dehesa, Theoretical Chemistry Accounts  \textbf{2016}, 135, 253.
	\bibitem{Lopezrosa2016} S. L\'opez-Rosa, M. Molina-Esp\'iritu, R.O.Esquivel, C. Soriano-Correa, J.S. Dehesa, ChemPhysChem \textbf{2016}, 17, 4003.
	\bibitem{Lopez1995} R. L\'opez-Ruiz, H. L. Mancini, and X. Calvet, Phys. Lett. A \textbf{1995}, 209, 321.
	\bibitem{Catalan2002} R. G. Catalán, J. Garay, R. López-Ruiz, Phys. Rev. E \textbf{2002}, 66, 011102.
	
	\bibitem{Romera2004} E. Romera, J. S. Dehesa. J. Chem. Phys. \textbf{2004}, 120, 8906.
	\bibitem{Angulo2008} J.C. Angulo, J. Antol\'in, K.D. Sen, Phys. Lett. A \textbf{2008}, 372, 670.
	\bibitem{Angulo2008a} J.C. Angulo, J. Antol\'in, J. Chem. Phys. \textbf{2008}, 128, 164109.
	
	\bibitem{Sabin2009}  \textit{The Theory of Quantum Confined Systems} (Eds: J. R. Sabin, E. Brandas, S. A. Cruz). Parts I and II. Adv. Quantum Chemistry, Academic Press, New York \textbf{2009}. 
	
	\bibitem{Sen2014} \textit{Electronic Structure of Quantum Confined Atoms and Molecules} (Ed: K.D. Sen), Springer, Berlin, \textbf{2014}. 
	\bibitem{Inglesfield2015} \textit{The Embedding Method for Electronic Structure} (Ed: J. E. Inglesfield) IOP Publishing, Bristol, \textbf{2015}.
	
	\bibitem{LeyKoo2018} E. Ley-Koo, Rev. Mex. Fis. \textbf{2018}, 64, 326.
	
	\bibitem{Cox1985} M. P.  Cox, G. Ertl, R. Imbihl, Phys. Rev. Lett. \textbf{1985}, 54, 1725.
	\bibitem{Jakubith1985} S. Jakubith, H. H. Rotermund, W. Engel, A. Von Oertzen, G. Ertl, Phys. Rev. Lett. \textbf{1990}, 65, 3013.
	\bibitem{Beta1985} C. Beta, M. G. Moula, A. S. Mikhailov, H. H Rotermund, G. Ertl, Phys. Rev. Lett. \textbf{2004}, 93, 188302.
	\bibitem{Li2007} S. S. Li, J. B. Xia, Phys. Lett. A \textbf{2007}, 366, 120.
	
	\bibitem{Harrison2005} P. Harrison, \textit{Quantum Wells, Wires and Dots: Theoretical and Computational Physics of Semiconductors Nanostructures}, Wiley-Interscience, New York \textbf{2005}.
	\bibitem{Munjal2018} D. Munjal, K.D. Sen, V. Prasad, J. Phys. Comm. \textbf{2018}, 2, 025002.
	
	
	
	
	
	
	\bibitem{Al-Hashimi2012} M. H. Al-Hashimi, U. J. Wiese, Ann. Phys. \textbf{2012}, 327, 2759.
	\bibitem{Marin} J. L. Marin and S. A. Cruz, J. Phys. B.: At. Mol. Opt. Phys. \textbf{1991}, 24, 2899, and references therein.
	
	
	\bibitem{Aquino2014} N. Aquino, A. Flores-Riveros, The confined hydrogen atom revisited. In \textit{Electronic Structure of Quantum Confined Atoms and Molecules} (Ed: K.D. Sen), Springer, Berlin, \textbf{2014}.
	
	\bibitem{Rojas2018} R. A. Rojas, N. Aquino, A. Flores-Riveros, Int. J. Quantum Chem. \textbf{2018}, 2018, e25584. 
	\bibitem{Gleisberg2000} F. Gleisberg, S. Wonneberger, U. Schlooder, C. Zimmermann. Phys. Rev. A \textbf{2000}, 62, 063602.
	\bibitem{Nascimento2011} W. Nascimento, F. V. Prudente, M. N. Guimaraes, A. M. Maniero, J. Phys. B.: At. Mol. Opt. Phys. \textbf{2011}, 44, 015003 .
	
	\bibitem{Anglin2002} J. R. Anglin, W. Ketterle. Nature \textbf{2002}, 416, 211.
	\bibitem{DeMarco1999} B. DeMarco , D. Jin. Sci. \textbf{1999}, 285, 1703.
	
	\bibitem{Lukin2018} H. Pichler, S. T. Wang, L. Zhou, S. Choi, M. D. Lukin, \textbf{2018}, arXiv:1809.04954.
	
	\bibitem{Yoder2019} T. J. Yoder, Phys. Rev. A \textbf{2019}, 99,  052333.
	
	\bibitem{Sen2005} K. D. Sen, J. Chem. Phys. \textbf{2005}, 123, 074110.
	\bibitem{Aquino2013} N. Aquino, A. Flores-Riveros, J.F. Rivas-Silva, Phys. Lett. A \textbf{2013}, 377, 2062.
	
	\bibitem{Nascimento2018} W. Nascimento, F. V. Prudente, Chem. Phys. Lett. \textbf{2018}, 691, 401.
	
	\bibitem{Jiao2017} L. G. Jiao, L. R. Zan, Y. Z. Zhang, Y. K. Ho, Int. J. Quantum Chem. \textbf{2017}, 117, e25375.
	
	\bibitem{Mukherjee2018}  N. Mukherjee, A. K. Roy, Eur. Phys. J. D \textbf{2018}, 72, 118.
	\bibitem{Mukherjee2018a} N. Mukherjee  A. K. Roy, Int. J. Quantum Chem. \textbf{2018}, 118, e25596.
	\bibitem{Mukherjee2018b} N. Mukherjee, S. Majumdar, A. K. Roy, Chem. Phys. Lett. \textbf{2018}, 691, 449.
	\bibitem{Garza2019} M. A. Mart\'inez-S\'anchez, R. Vargas, J. Garza, Quantum Rep. \textbf{2019}, 1, 208.
	
	\bibitem{Wu2020} L. Wu, S. Zhang, B. Li,  Phys. Lett. A \textbf{2020}, 384, 126033.
	
	\bibitem{Majumdar2017} S. Majumdar, N. Mukherjee, A. K. Roy, Chem. Phys. Lett. \textbf{2017}, 687, 322.
	\bibitem{Yang1991} X. L. Yang, S. H. Guo, F. T. Chan, K. W. Wong, W. Y. Ching, Phys. Rev. A \textbf{1991}, 43, 1186.
	\bibitem{Aquino1998a} N. Aquino, E.  Castaño, Rev. Mex. F\'{\i}s. \textbf{2005}, 51, 126.
	\bibitem{Aquino1998} N. Aquino, E. Casta\~{n}o, Rev. Mex. F\'{\i}s. \textbf{1998}, 44, 628. 
	\bibitem{Rojas2019} R.A. Rojas, N. Aquino, Rev. Mex. F\'{\i}s. \textbf{2019}, 65, 116. 
	
	\bibitem{Aquino2005} N. Aquino, G. Campoy, A. Flores-Riveros, Int. J. Quantum. Chem. \textbf{2005}, 103, 267.
	\bibitem{Chaos2005} L. Chaos-Cador, E. Ley-Koo, Int. J. Quantum. Chem. \textbf{2005}, 103, 369.
	\bibitem{Estanon2020} C.R. Estañón, N. Aquino, D. Puertas-Centeno, J.S. Dehesa, Int. J. Quantum Chem. \textbf{2020}, e26192.
	\bibitem{Yanez1994} R. J. Yáñez, W. Van Assche, J. S. Dehesa, Phys. Rev. A \textbf{1994}, 50, 3065.
	\bibitem{Dehesa2006} J. S. Dehesa, S. L\'opez-Rosa, B. Olmos, R. Yáñez, J. Math. Phys. \textbf{2006}, 47, 052104.
	
	\bibitem{Romera2006} E. Romera, P. S\'anchez-Moreno, J.S. Dehesa, J. Math. Phys. \textbf{2006}, 47, 103504.
	
	\bibitem{Aptekarev2010} A-I. Aptekarev, J. S. Dehesa, A. Mart\'inez-Finkelshtein, R. Yáñez, J. Phys. A \textbf{2010}, 43, 145204.
	\bibitem{Dehesa2012} J.S. Dehesa, A.R. Plastino, P. S\'anchez-Moreno, C. Vignat, Appl. Math, Lett. \textbf{2012}, 25, 1689.
	\bibitem{Dembo1991} A. Dembo, T.M. Cover, J.A. Thomas, IEEE Trans. Inform. Theor. \textbf{1991}, 37, 1501.
	\bibitem{Dehesa2007a} J.S. Dehesa, R. Gonz\'alez-F\'erez, P. S\'anchez-Moreno, J. Phys. A \textbf{2007}, 40, 1845.
	\bibitem{Sanchez-Moreno2011} P. Sánchez-Moreno, A. R. Plastino, J. S. Dehesa, J. Phys A \textbf{2011}, 44, 065301.
	\bibitem{Mukherjee2018c} N. Mukherjee, A.  Roy, A. K. Roy, Ann. Phys. \textbf{2018}, 398, 190.
	\bibitem{Yu2017} R.M. Yu, L.R. Zan, L.G. Jiao, Y.K. Ho, Few Body Syst. \textbf{2017}, 58, 152.
	
	\bibitem{Ping2019} J. Ping, H.S. Zong, ArXiv:1902.05355v1 [physics.atom-ph] (14 February \textbf{2019})
	\bibitem{Assche2000} W. van Assche, R. J. Yáñez, R. González-Férez and J. S. Dehesa, J.Math.Phys. {\bf 2000}, 41, 6600.	
	\bibitem{Adkins2019} G.S. Adkins, M. F. Alam, C. Larison, and R. Sun, arXiv:1908.02324v2 [quant-ph] 9 Dec \textbf{2019}.
	
	\bibitem{Delbourgo2009} R. Delbourgo and D. Elliott, J. Math. Phys. \textbf{2009}, 50, 062107.
\end{thebibliography}
\end{document}